\documentclass[10pt]{article}
\usepackage{amsmath}
\usepackage{amssymb}
\usepackage{graphicx}
\usepackage{cite}
\usepackage{color} 
\usepackage[labelfont=bf,labelsep=period,justification=raggedright]{caption}
\makeatletter
\renewcommand{\@biblabel}[1]{\quad#1.}
\makeatother

\date{}

\begin{document}

% Title must be 150 characters or less
\begin{flushleft}
{\Large
\textbf{Phase transition in p53 states induced by glucose}
}
\bigskip
\\
Md. Jahoor Alam$^{1,2}$ and R.K. Brojen Singh$^{2*}$.
\\
\bigskip
$^1$College of Applied Medical Sciences, University of Ha'il, Ha'il-2440, Saudi Arabia.\\ $^2$School of Computational and Integrative Sciences, Jawaharlal Nehru University, New Delhi-110067, India.
\\
\bigskip
$\ast$ Corresponding author, E-mail:  R.K. Brojen Singh - brojen@jnu.ac.in, \\

\end{flushleft}

\section*{Abstract}
We present p53-MDM2-Glucose model to study spatio-temporal properties of the system induced by glucose. The variation in glucose concentration level triggers the system at different states, namely, oscillation death (stabilized), sustain and damped oscillations which correspond to various cellular states. The transition of these states induced by glucose is phase transition like behaviour. We also found that the intrinsic noise in stochastic system helps the system to stabilize more effectively. Further, the amplitude of $p53$ dynamics with the variation of glucose concentration level follows power law behaviour, $A_s(k)\sim k^\gamma$, where, $\gamma$ is a constant.
\bigskip

{\bf Keywords:} Glucose, p53, DNA damage, Oscillating states.

\subsection*{Introduction}
Oscillations are inherent and inbuilt in living systems due to various fundamental molecular processes and coordinate basic biological functions and their mechanisms \cite{win} to self-organize the complicated life processes \cite{fei}. The origin of these oscillations could be due to various control or feedback mechanisms and set of non-linearities which describe complicated non-linear activities in the system \cite{bak}.
$p53$ is one of most important proteins in cellular system which contribute to the maintenance of the genomic integrity \cite{alb}, involves in various cellular activities, such as, cell cycle arrest, DNA repair, apoptosis and other cellular functions \cite{smi,oli,mid}, and exhibits oscillatory behaviour by interacting $MDM2$ via feedback mechanism \cite{mid}. Regulating $p53$, variety of cellular stresses, for example, global DNA damage that cause abnormal or cancerous cells and repairing of stress-induced DNA damage, can be controlled \cite{smi,ler}. Even though, $p53$ is functionally inhibited in normal cells \cite {oli,fin}, the $MDM2$, which is a negative regulator of $p53$, can activate $p53$ to induce stress in the cell \cite{mid,bra,chen}.

The oscillatory behaviour exhibited by $p53$ due to negative feedback mechanism with $MDM2$ can be regulated and induce stress by other molecular activities, such as, $Ca^{2+}$, $NO$ \cite{ala1}, $MTBP$ \cite{ala} etc. and excess stress may lead to apoptosis \cite{chen,ala,hau}. However, if the stress induced by various stress inducers ($NO$, reactive oxygen synthase ($ROS$), $ARF$ etc) is weak, on removing stress $p53$ activation may come back to its normal functioning \cite{thu} which can be done by $MDM2$ by enhancing the ubiquitination of $p53$, as a result of which $p53$ level degrades \cite{hau,kub}. 

However, there are many open issues regarding the role of glucose in cellular activities, for example, switching of normal to stress states via glucose level in the cell, the way how excess stress caused by glucose lead the cell to apoptosis, the possibilities to estimate the critical concentration of glucose that cause the state transitions, the way how glucose effects p53 dynamics etc. We, in this work, study the impact of glucose on $ROS$ activation which cause DNA damage in p53-MDM2-Glucose network to understand different states in the system and their transitions. The work is organized as follows. We describe p53-MDM2-Glucose model and its molecular interaction with numerical techniques in section 2. The numerical results are presented with discussions in section 3 and some conclusions are drawn based on the results we obtained in section 4.

\subsection*{Mathematical model of $p53-MDM2-Glucose$ network}

The model we consider is the extension of $p53-MDM2$ network \cite{pro} induced by two new and important molecules namely stress inducer glucose and ROS, which damages DNA (Fig. 1). In the model, p53 acts as a transcription factor and helps in the transcription of $MDM2\_mRNA$ via $MDM2~gene$. $MDM2\_mRNA$ then synthesizes MDM2 protein through the translation process. p53 interacts with MDM2 by enhancing its degradation through its E3 ubiquitin ligase activity \cite{pro} and maintains low concentration level at normal condition \cite{bra}. We then consider glucose metabolism inside cell which leads to production of ROS \cite{rob} and high ROS concentration triggers the DNA damage \cite{car,ama} inducing stress to the system \cite{rob2,dan}. The DNA damage activates ARF and this activated ARF interacts with MDM2 forming $ARF\_MDM2$ complex \cite{khan,she}. This activity of ARF blocks the MDM2 E3 ubiquitin ligase activity which promotes MDM2 degradation \cite{she}. 
Our model (Fig. 1) consists of nine molecular species listed in Table 1 which undergo the reaction channels listed in Table 2. 

\begin{center}
{{\bf Table 1} List of molecular species} 
    \begin{tabular}{ | l | p{3.5cm} | p{4.0cm} | p{2.0cm} |}
       \hline \multicolumn{4}{}{} \\ \hline
        {\bf S.No} & {\bf Molecular Species} & {\bf Description} & {\bf Notation} \\ \hline
        1. & p53 & Unbound p53 protein & $x_1$ \\ \hline
        2. & MDM2 & Unbound MDM2 protein & $x_2$ \\ \hline
        3. & $p53\_MDM2$ & p53/MDM2 complex & $x_3$ \\ \hline
        4. & $MDM2\_mRNA$ & MDM2 messenger RNA & $x_4$ \\ \hline
        5. & Glucose & Unbound Glucose & $x_5$ \\ \hline
        6. & ROS & Unbound ROS & $x_6$ \\ \hline
        7. & $Dam\_DNA$ & Damage DNA & $x_7$ \\ \hline
        8. & ARF & ARF protein & $x_8$ \\ \hline
        9. & $ARF\_MDM2$ & ARF/MDM2 complex & $x_9$ \\ \hline
        
\end{tabular}
\end{center}

\begin{table*}
\begin{center}
{\bf Table 2 List of biochemical reaction, Kinetic Law and their rate constant} 
 \begin{tabular}{|l|p{2.5cm}|p{3.0cm}|p{1.5cm}|p{2.5cm}|p{3cm}|}
 \hline \multicolumn{6}{}{} \\ \hline

        {\bf S.No} & {\bf Reaction channel} & {\bf Description} & {\bf Kinetic Law} & {\bf Values of rate constant}&{\bf References}\\ \hline
        1  & $x_4\stackrel{k_1}{\longrightarrow}x_4+x_2$ & MDM2 translation &  $k_1 x_3 $ & $4.95\times10^{-4} sec^{-1}$ & \cite{fin,ala,pro}. \\ \hline
        2  & $x_1\stackrel{k_2}{\longrightarrow}x_1+x_4$ & Synthesis of $MDM2\_mRNA$ & $k_2 x_1 $ & $1.0\times 10^{-4} sec^{-1}$ & \cite{fin,ala,pro}.\\ \hline
        3  & $x_4\stackrel{k_3}{\longrightarrow}\phi$ & Degradation of $MDM2\_mRNA$ & $k_3 x_3 $ & $1.0\times 10^{-4} sec^{-1}$ & \cite{fin,ala,pro}.\\ \hline
        4  & $x_2\stackrel{k_4}{\longrightarrow}\phi$ & Degradation of MDM2 & $k_4 x_2 $ & $4.33\times 10^{-4} sec{-1}$ & \cite{fin,ala,pro}. \\ \hline
        5  & $\phi\stackrel{k_5}{\longrightarrow}x_1$ & Synthesis of p53 &  $k_5$ & $0.78 mol sec^{-1}$ & \cite{fin,ala,pro}.\\ \hline
        6  & $x_3\stackrel{k_6}{\longrightarrow}x_2$ & Decay of p53 &  $ k_6 x_3 $ & $8.25\times 10^{-4} sec^{-1}$ & \cite{fin,ala,pro}. \\ \hline
        7  & $x_1+x_2\stackrel{k_7}{\longrightarrow}x_3$ & Synthesis of p53\_MDM2 complex &  $k_7 x_1 x_2 $ & $11.55\times 10^{-4} mol^{-1}sec^{-1}$ & \cite{fin,ala,pro}. \\ \hline
        8  & $x_3\stackrel{k_8}{\longrightarrow}x_1+x_2$ & Dissociation of p53\_MDM2 complex &  $k_8 x_3 $ &$11.55\times 10^{-6} sec^{-1}$ & \cite{fin,ala,pro}. \\ \hline 
        9  & $\phi\stackrel{k_{9}}{\longrightarrow}x_5$ & Creation of Glucose & $k_9$ & $k$ [$mol^{-1}sec^{-1}$]& Assuming the concentration level of glucose vary within the cell. \\ \hline
        10 & $x_5\stackrel{k_{10}}{\longrightarrow}x_6$ & Synthesis of ROS  & $ k_{10} x_5 $ & $2\times10^{-3} mol^{-1}sec^{-1}$ & Assuming ros production due to the glucose molecule.\\ \hline
        11 & $x_6\stackrel{k_{11}}{\longrightarrow}x_7$ & DNA damage & $k_{11} x_6 $ & $5\times10^{-4} sec^{-1}$ & \cite{car,wis,mar}.\\ \hline
        12 & $x_7\stackrel{k_{12}}{\longrightarrow}\phi$ & DNA repair & $k_{12} x_7 $ & $2\times10^{-5} sec^{-1}$ & \cite{fin,pro}.\\ \hline
        13 & $x_7\stackrel{k_{13}}{\longrightarrow}x_8$ & ARF activation & $k_{13} x_7 $ & $3.3\times10^{-5} sec^{-1}$ & \cite{fin,pro}.\\ \hline
        14 & $x_2+X_8\stackrel{k_{14}}{\longrightarrow}x_9$ & Synthesis of ARF/MDM2 complex & $k_{14} x_2 x_8 $  & $1\times10^{-2} mol^{-1}sec^{-1}$ & \cite{fin,pro}.\\ \hline
        15 & $x_9\stackrel{k_{15}}{\longrightarrow}x_8$ & ARF dependent MDM2 degradation & $k_{15} x_9 $  & $1\times10^{-3} sec^{-1}$ & \cite{fin,pro}. \\ \hline
        16 & $x_8\stackrel{k_{16}}{\longrightarrow}\phi$ & Degradation of ARF & $k_{16} x_8 $  & $1\times10^{-3} sec^{-1}$ & \cite{fin,pro}.\\ \hline
        17 & $x_5\stackrel{k_{17}}{\longrightarrow}\phi$ & Degradation of Glucose & $k_{17} x_5 $  & $1\times10^{-4} sec^{-1}$ & Due to half life of the molecule.\\ \hline

\end{tabular}
\end{center}
\end{table*}

Cellular processes are basically noise induced stochastic processes \cite{ark} and this noise could be intrinsic due to random molecular interaction in the system \cite{gil1} as well as extrinsic due to fluctuations of physical variables surrounding the system \cite{elo}. The intrinsic molecular interactions in our model system undergo seventeen reaction channels (Table 2) and can be well explained in stochastic manner\cite{gil1,gil2,gil3} as follows. 

Consider a configurational state of our model system at any instant of time $t$ is defined by a state vector, $\vec X(t)=[X_1(t),X_2(t),\dots,X_N(t)]^T$, where, $\{X_i\}$ is the set of variables corresponding to populations of the molecular species (Table 1), $N=9$, and $T$ is the transpose of the vector. The time evolution of the configurational probability $P(\vec X;t)$ to have a transition from one configurational state $\vec{X}$ to another state $\vec{X(t)^\prime}$ during the time interval $[t, t+dt]$ is given by the following Master equation,
%\begin{widetext}
\begin{eqnarray}
\label{mas}
\frac{\partial P(\vec X(t),t)}{\partial t}&=&\sum_{\{\vec X\}}P(\vec X(t),t)\Gamma_{\vec X\rightarrow \vec X^\prime}-\sum_{\{\vec X^\prime\}}P(\vec X^\prime,t)\Gamma_{\vec X^\prime\rightarrow \vec X}
\end{eqnarray} 
%\end{widetext}
where, $\{\Gamma\}$ is the set of transition rates from one state to another. Since solving equation (\ref{mas}) is very difficult for our system, we follow Gillespie \cite{gil2} to simplify it to Chemical Langevin equations (CLE) as follows. The number of reactions fired during the time interval $[t,t+\Delta t]$ is a variable $B(a)$ which depends on propensity functions ($a$) of the reactions, and one can impose two important realistic approximations in the large population limit to arrive at CLE. First in the limit, $\Delta t\rightarrow 0$, the values of $a$ will remain constant during $[t,t+\Delta t]$, which allows $B$ to be replaced by statistically independent Poisson random variable. The second approximation is $\Delta t\rightarrow\infty$ which allows to approximate Poisson random variable by a normal variable, $G$ with the same mean and variance. Now linearising $G$, we reach the following CLE for our model,
%\begin{widetext}
\begin{eqnarray}
\label{cle}
\frac{dx_1}{dt}&=&k_5-k_7x_1x_2+k_8x_3+\frac{1}{\sqrt{V}}\bigg[\sqrt{k_5}\xi_1-\sqrt{k_7x_1x_2}\xi_2+\sqrt{k_8x_4}\xi_3\bigg]\\
\frac{dx_2}{dt}&=&k_1x_4-k_4x_2+k_6x_3-k_7x_1x_2+k_8x_3-k_{14}x_2x_8\nonumber \\
&&+\frac{1}{\sqrt{V}}\bigg[\sqrt{k_1x_4}\xi_4-\sqrt{k_4x_2}\xi_5+\sqrt{k_6x_4}\xi_6-\sqrt{k_7x_1x_2}\xi_7 \nonumber \\
&&+\sqrt{k_8x_3}\xi_8-\sqrt{k_{14}x_2x_8}\xi_9\bigg]\\
\frac{dx_3}{dt}&=&-k_6x_3+k_7x_1x_2-k_8x_3\nonumber\\
&&+\frac{1}{\sqrt{V}}\bigg[-\sqrt{k_{6}x_3}\xi_{10}+\sqrt{k_{7}x_1x_2}\xi_{11}-\sqrt{k_8x_3\xi_{12}}\bigg]\\
\frac{dx_4}{dt}&=&k_2x_1-k_3x_4+\frac{1}{\sqrt{V}}\bigg[\sqrt{k_{2}x_1}\xi_{13}-\sqrt{k_{3}x_4}\xi_{14}\bigg]\\
\frac{dx_5}{dt}&=&k_9-k_{10}x_5-k_{17}x_5+\frac{1}{\sqrt{V}}\bigg[\sqrt{k_9}\xi_{15}-\sqrt{k_{10}x_5}\xi_{16}-\sqrt{k_{17}x_5}\xi_{17}\bigg]\\
\frac{dx_6}{dt}&=&k_{10}x_5-k_{11}x_6+\frac{1}{\sqrt{V}}\bigg[\sqrt{k_{10}x_5}\xi_{18}-\sqrt{k_{11}x_6}\xi_{19}\bigg]\\
\frac{dx_7}{dt}&=&k_{11}x_6-k_{12}x_7+\frac{1}{\sqrt{V}}\bigg[\sqrt{k_{11}x_6}\xi_{20}-\sqrt{k_{12}x_7}\xi_{21}\bigg]\\
\frac{dx_8}{dt}&=&k_{13}x_7-k_{14}x_2x_8+k_{15}x_9-k_{16}x_8+\frac{1}{\sqrt{V}}\bigg[\sqrt{k_{13}x_7}\xi_{22}\nonumber\\
&&-\sqrt{k_{14}x_2x_8}\xi_{23}+\sqrt{k_{15}x_9}\xi_{24}+\sqrt{k_{16}x_8}\xi_{25}\bigg]\\
\frac{dx_9}{dt}&=&k_{14}x_2x_8-k_{15}x_9+\frac{1}{\sqrt{V}}\bigg[\sqrt{k_{14}x_2x_8}\xi_{26}-\sqrt{k_{15}x_9}\xi_{27}\bigg]
\end{eqnarray} 
%\end{widetext}
where, $\{\xi_i\}$ are random parameters which is given by, $\xi_i$ = $lim_{dt\rightarrow 0}N_i(0,1)/\sqrt{dt}$ and satisfy $\xi_i(t)\xi_j(t^\prime)$ = $\delta_{ij}\delta(t-t^\prime)$. When $V\rightarrow\infty$ or $\{\xi\}\rightarrow 0$ the set of CLEs recover deterministic equations.
\begin{figure*}
\begin{center}
\includegraphics[height=400 pt]{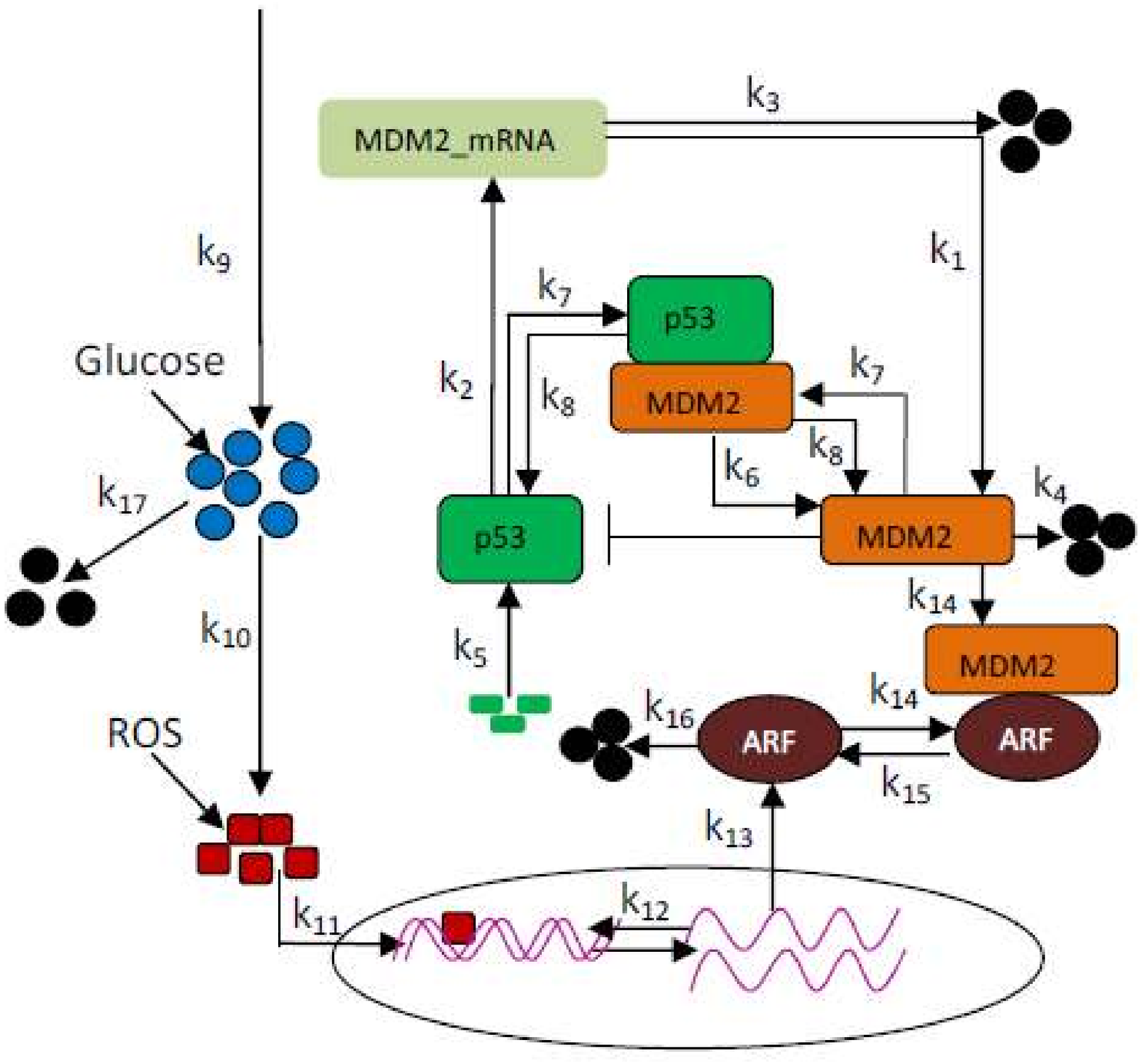}
\caption{The schematic diagram of p53-MDM2-Glucose interaction network which involves various feedback loops induced by ARF protein and regulated indirectly by Glucose.}
\end{center}
\end{figure*}

The dynamical behaviour of the system can be well studied by simulating the set of CLEs given by equations (2)-(10) or corresponding deterministic differential equations by using standard 4th order Runge-Kutta method \cite{pre} for numerical integration of the set of stochastic or ordinary differential equations.

We use the stochastic simulation algorithm due to Gillespie \cite{gil1} to calculate the time evolution of state vector of the system by simulating the reaction sets given in Table 2 with the parameter values. The algorithm is based on proposed joint probability density function $P(\tau,\mu)=T(\tau)R(\mu)$ to allow a transition to occur with time increment $\tau~(=[0,\infty])$ by finding the probability densities, $T(\tau)$ at which a particular reaction is fired and $R(\mu)$ which identifies the particular reaction $\mu=[1,2,..,M]$ fired at that time. From this hypothesis, $\tau$ and $\mu$ can be estimated computationally from the relations $\tau=-ln[T(\tau)]/a_o$ and $\omega_\mu=\omega_oR(\mu)$ by defining two random numbers $r_1$ and $r_2$ for $T(\tau)$ and $R(\mu)$ respectively and here $a_o=\sum_{i=1}^Ma_i$.
\begin{figure*}
\begin{center}
\includegraphics[height=350 pt]{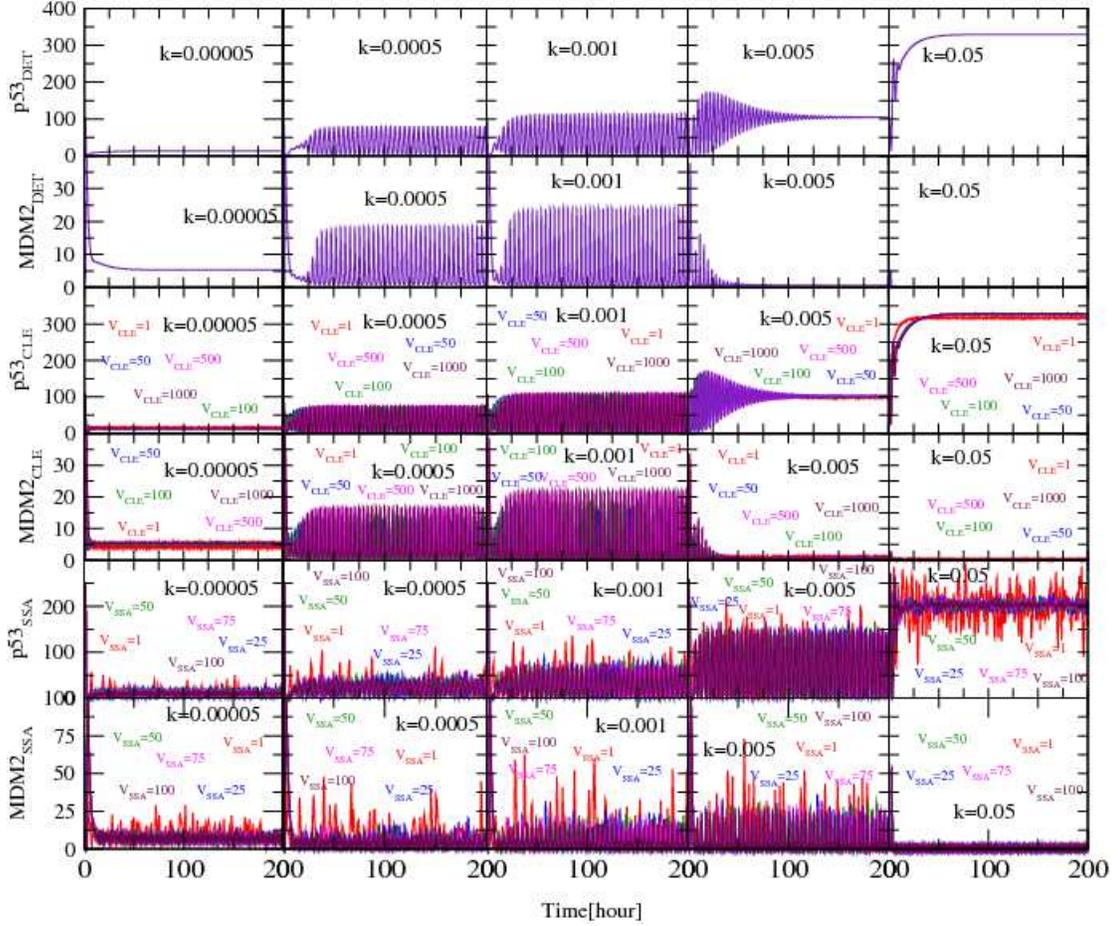}
\caption{Plots of a comparative behaviour of $p53$ and $MDM2$ as a function of time in hours (for 0 to 200 hours) due to the effect of rate constant of Glucose creation, for different values of $k_9$=$k$=0.00005 ,0.0005, 0.001, 0.005, 0.05 respectively, (i) Deterministic case (the first upper two row panels), (2) Chemical Langevin equation (CLE) (the third and fourth row panels) and (3) Stochastic simulation algorithm (SSA) (the fifth and sixth row panels). For CLE we have taken five values of system sizes i.e. V=1, 50, 100, 500, 1000; and for SSA we have taken V=1, 25, 50, 75 and 100.}
\end{center}
\end{figure*}

\subsection*{Results and discussion}

The dynamical behaviours of $p53$ and various states of the system induced by glucose concentration levels in it are studied using three different computational techniques, namely, deterministic by solving equations (2)-(10) where $\{\xi\}\rightarrow 0$, CLE by solving equations (2)-(10) and stochastic simulation algorithm by simulating the set of reactions listed in Table 2. The parameter values needed for the simulation are also given in Table 2.
\begin{figure*}
\begin{center}
\includegraphics[height=350 pt]{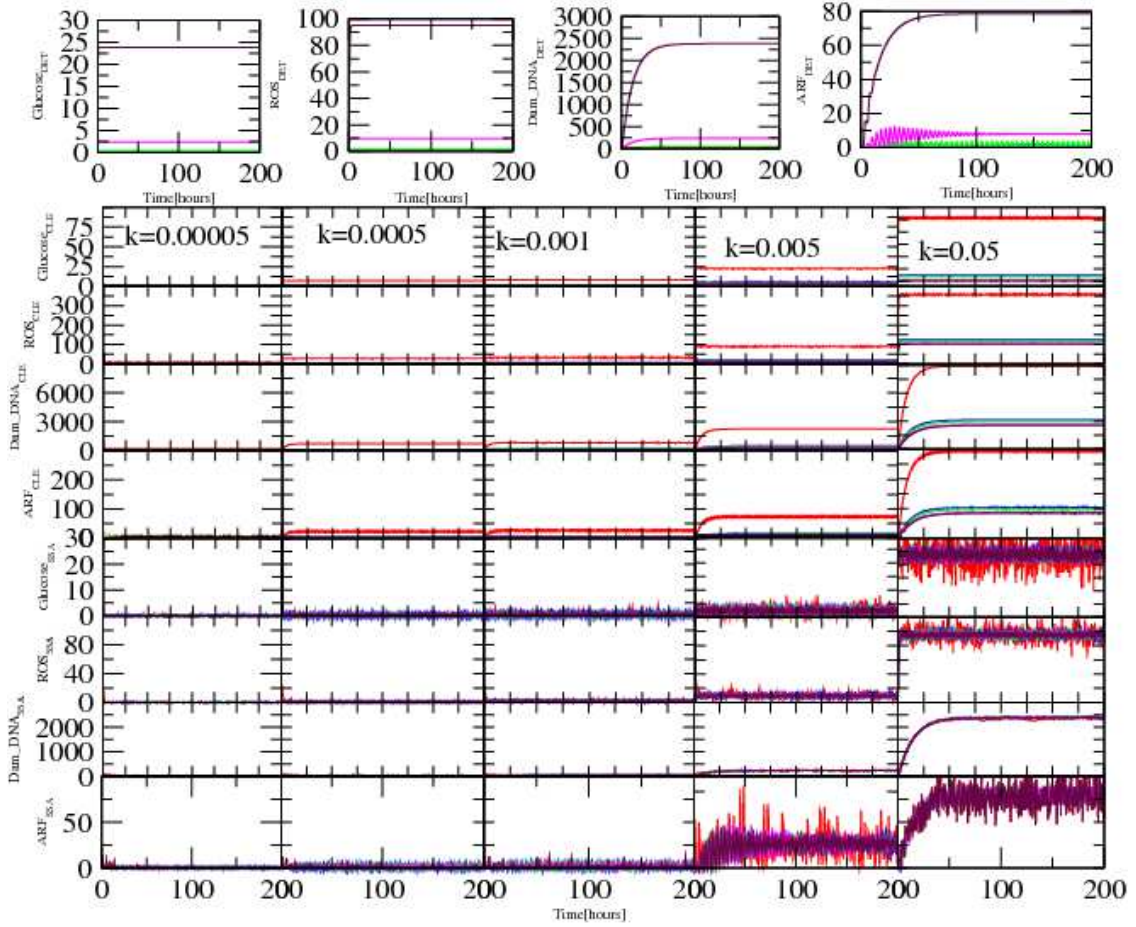}
\caption{Similar plot as shown in figure 2,  of comparative behaviour of $Glucose$, $ROS$, $Damaged_DNA$ and $ARF$ as a function of time in hours (for 0 to 200 hours) due to the effect of rate constant of Glucose creation, for different values of $k_9$=$k$=0.00005 ,0.0005, 0.001, 0.005, 0.05 respectively, (i) Deterministic case (the first upper row panels), (2) Chemical Langevin equation (CLE) (the second, third, fourth and fifth row panels) and (3) Stochastic simulation algorithm (SSA) (the sixth, seventh, eighth and nineth row panels). For CLE we have taken five values of system sizes i.e. V=1, 50, 100, 500, 1000; and for SSA we have taken V=1, 25, 50, 75 and 100. (for different colours code see figure 2).}
\end{center}
\end{figure*}

\subsubsection*{Transition of oscillating states: biological rules}

The glucose concentration level (proportional to the rate $k$ which is the rate of creation of glucose given in reaction number 9 in Table 2) in the system, due to both synthesized by cellular processes and diffused in from extracellular medium, drives the system to different oscillating states (Fig. 2). The $p53$ dynamics maintains stabilized state with minimum $p53$ level at low values of $k$ ($k\langle 0.00007$) and this state is not much influenced by intrinsic noise in the system except some random fluctuation about the stabilized state (Fig. 2 last four panels of first column). The two dimensional plots of $p53$ and $p53\_MDM2$ in this situation shows fixed point oscillation (oscillation with amplitude zero)  (Fig. 4). This stabilized state of $p53$ reflects the normal condition in cellular systems where $p53$ is maintained minimum level\cite{ala,gev}.
\begin{figure*}
\begin{center}
\includegraphics[height=200 pt, width=400 pt]{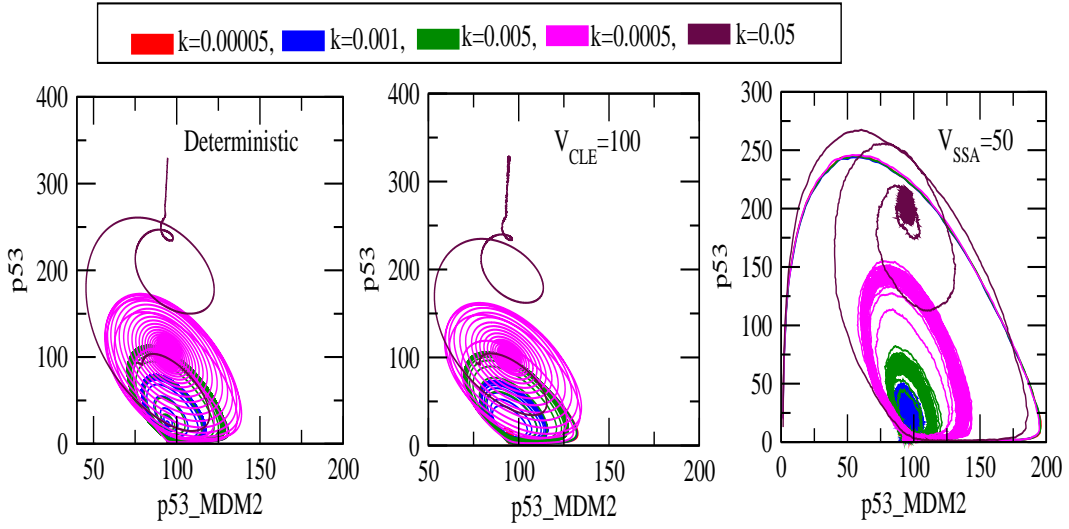}
\caption{Two dimensional plots of p53 for different values of $k_9$ in deterministic, CLE (V=100) and SSA (V=50).}
\end{center}
\end{figure*}

The moderate concentration level of glucose in the system ($0.0005\langle k\langle 0.001$) drives the $p53$ behaviour to sustain oscillation state with increasing amplitudes as $k$ increases (Fig. 2 second and third columns). In this condition, two dimensional plots of $p53$ and $p53\_MDM2$ show broaden limit cycle (Fig. 4). The random fluctuation in $p53$ dynamics is minimized in deterministic system and start showing up significantly in CLE and SSA where the fluctuations increases as noise strength increases (as $V$ decreases). The continuous and periodic changes in the $p53$ concentration level show the active participation of $p53$ in the molecular interaction described in Table 2 which involve creation and degradation of $p53$ and other molecular species in the system. This sustain oscillation state, therefore, may correspond to activated or stress state in cellular system induced by glucose.

The further increase in the glucose concentration level in the system ($0.005\langle k\langle 0.008$) drives $p53$ dynamics from sustain to damped oscillation state and then to stabilized state (Fig. 2 fourth column). In this situation, two dimensional plots of $p53$ and $p53\_MDM2$ show spiral cycle towards fixed point (Fig. 4). Even though this is a clear damped state in deterministic system and CLE (noise strength is small), in SSA (noise strength is comparatively large) the state is still sustain oscilation state with large amplitude. This indicates that noise can also be taken as a parameter which can induce stress to the system. 

However, if the glucose concentration level is high enough then the $p53$ dynamics goes to the stabilized state again with high population (Fig. 2 last column). Similary, the comparative behaviour of $Glucose$, $ROS$, $Damaged\_DNA$ and $ARF$ is shown in fig. 3, as a function of time in hours (for 0 to 200 hours) due to the effect of rate constant of Glucose creation, for different values of $k_9$=$k$=0.00005 ,0.0005, 0.001, 0.005, 0.05 respectively, (i) Deterministic case (the first upper row panels), (2) Chemical Langevin equation (CLE) (the second, third, fourth and fifth row panels) and (3) Stochastic simulation algorithm (SSA) (the sixth, seventh, eighth and nineth row panels). For CLE we have taken five values of system sizes i.e. V=1, 50, 100, 500, 1000; and for SSA we have taken V=1, 25, 50, 75 and 100. (for different colours code see figure 2). The result suggests that as the rate constant of glucose (which corresponds to glucose concentration in the system) increaeses the stress within the systems first increases then finally acheived an steady state at high value of rate constant.

In this condition, two dimensional plots of $p53$ and $p53\_MDM2$ show fixed point oscillation again (Fig. 4). In stochastic system, there is still oscillating behaviour at large strength of noise ($V=1$) which reveals that still noise tries to resist the system to go to stabilized state (Fig. 2 last column fifth panel). This second stabilized state may correspond to apoptotic state because increasing glucose level the $p53$ dynamics will remain stabilized forever.

The time of activation $T_s$ for p53 temporal dynamics, which can be defined as the time below which the dynamics show damped oscillation and above which it shows oscillation death or fixed point oscillation, is calculated for different $k$ values (Fig. 6) for deterministic, CLE and SSA. This phase diagram shows various regimes of oscillation states and their switching boundaries. The error bars in each curve are due to average over 30 ensembles. Further, we calculated amplitudes of the $p53$ dynamics ($A_s$) as a function of $k$ (Fig. 6 right column) which obeys power law behaviour with $k$ in large $k$ regime i.e. $A_s(k)\sim k^{\gamma}$. All the curves of different V are found to be within the error bars showing the similar behaviour of the system as a function of V.
\begin{figure*}
\begin{center}
\includegraphics[height=200 pt, width=400 pt]{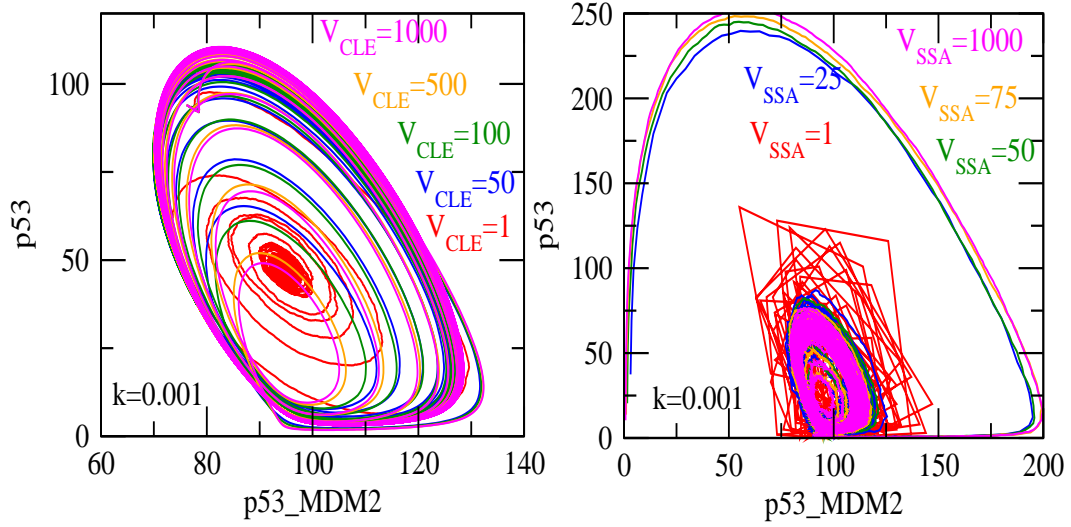}
\caption{Two dimensional plots of $p53$ for different values of $V$ for fixed value of $k$: (i) CLE for $V=1, 50, 100, 500, 1000$ and (ii) SSA for the same values of $V$.}
\end{center}
\end{figure*}

\subsubsection*{Noise to stabilize the system}

Noise has constructive role in regulating the system by trying to keep the system at normal condition (stabilized condition), on removing the noise (as $V$ increases) the system goes to activated or stress state (Fig. 5). Further, noise also tries to save the stress state from going to apoptotic state (Fig. 2 fifth panels of fourth and fifth columns). It is also evident from the curves in Fig. 6 that as $V$ decreases the curves shifts on the right hand side showing (i) helping the system to maintain in normal state by keeping away from stress, indicated by larger area occupied towards oscillation death regime as $V$ decreases (left hand side of sustain oscillation in Fig. 6 left column) (ii) allowing the system to save from apoptosis by slowing down to reach stabilized state (larger area occupied in oscillation death regime in right hand side of sustain oscillation in Fig. 6 left column).

\subsubsection*{Stability solution}

The stabilized solution of $p53$ and $MDM2$ can be obtained from the stationary conditions of the set of equations (2)-(10) by putting $\frac{d}{dt}{\bf x}=0$ and $\xi_i\rightarrow 0, \forall i$. Solving for $x_1^*$ from the nine stationary equations we obtain the following equation,
\begin{eqnarray}
\label{x1}
x_1^*\sim A\sqrt{x_5^*}\left(1+\frac{B}{x_5^*}\right)
\end{eqnarray}
where, $A=\sqrt{\frac{k_3k_5k_{10}k_{13}k_{14}}{k_2k_7k_{12}k_{16}}\left(1+\frac{k_8}{k_6}\right)}$ and $B=\frac{k_4k_{12}k_{16}}{2k_{10}k_{13}k_{14}}$. The equation (\ref{x1}) shows that for large values of glucose concentration $x_5^*$ (stabilized condition for large glucose concentration), $B/x_5^*\rightarrow 0$ which gives $x_1^*\propto \sqrt{x_5^*}$. For small values of glucose concentration, $1+B/x_5^*\sim B/x_5^*$, and we found that $x_1^*\propto\frac{1}{\sqrt{x_5^*}}$ maintaining low p53 concentration level.

Similarly, the stabilized solution for $MDM2$ ($x_2^*$) is given by,
\begin{eqnarray}
\label{x2}
x_2^*\sim C\frac{\sqrt{x_5^*}}{B+x_5^*}
\end{eqnarray}
where, $C=\sqrt{\frac{k_2k_5k_{12}k_{16}}{k_3k_7k_{10}k_{13}k_{14}}\left(1+\frac{k_8}{k_6}\right)}$. At large glucose concentration level, $1+B/x_5^*\rightarrow 1$ and equation (\ref{x2}) gives, $x_2^*\propto\frac{1}{\sqrt{x_5^*}}$. Further, for low glucose concentration level $1+x_5^*/B\rightarrow 1$ such that $x_2^*\propto x_5^*$ showing Mdm2 level is at large as compared to higher glucose concentration level.
\begin{figure*}
\begin{center}
\includegraphics[height=280 pt]{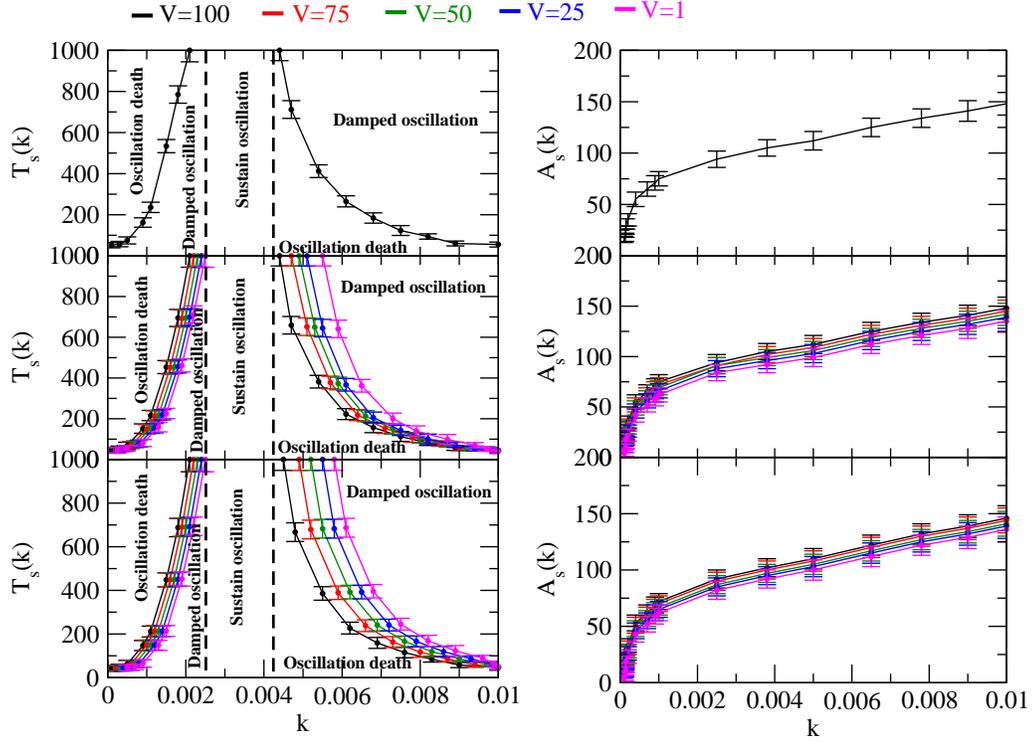}
\caption{The phase diagram for different $V$ in $(T_s-k)$ plane and $(A_s-k)$ are presented for (i) deterministic ($V\rightarrow\infty$), (ii) CLE (V=1, 25, 50, 75, 100) and (iii) SSA (V=1, 25, 50, 75, 100).}
\end{center}
\end{figure*}

\subsection*{Conclusion} 
We investigated transition of various oscillatory states in p53-MDM2-Glucose model induced by glucose \cite{ama}. Our simulation results in three approaches namely deterministic, CLE and SSA show three distinct states, namely, oscillation death, damped and sustained oscillatiory states, and a clear transition among these states induced by glucose concentration level. This transition could be the signature of transition of $p53$ and $MDM2$ states from normal to stress, stress to apoptotic state induced by glucose. 

Intrinsic noise associated with the system dynamics helps the system to maintain its stabilized state and further helps to protect from apoptosis. However, there are other several points to be studied further on the roles of noise, finding critical strength of noise below which the role of noise is constructive, role of external noise in maintaining orderness in non-equilibrium and non-linear physical, chemical and biological systems.

\subsection*{Acknowledgments}
We thank Prof. Pankaj Sharan and Prof. R. Ramaswamy for stimulating comments and discussions in carrying out this work. This work is financially supported by UPE-II, under Project no. 101.

\end{document}